\documentclass{sig-alternate-2013}
\newtheorem{Example}{\it Example}[section]

\newenvironment{proof-of}[1]{{\em Proof of #1:}}{}

\usepackage{color}
\usepackage{graphicx}
\usepackage{etoolbox}
\usepackage{textcomp}

\newfont{\mycrnotice}{ptmr8t at 7pt}
\newfont{\myconfname}{ptmri8t at 7pt}
\let\crnotice\mycrnotice%
\let\confname\myconfname%

\permission{Permission to make digital or hard copies of part or all of this work for personal or classroom use is granted without fee provided that copies are not made or distributed for profit or commercial advantage and that copies bear this notice and the full citation on the first page. Copyrights for third-party components of this work must be honored. For all other uses, contact the Owner/Author. 
Copyright is held by the owner/author(s).}
\conferenceinfo{ACM DEV 2015,}{December 1--2, 2015, London, United Kingdom.} 
\copyrightetc{ACM \the\acmcopyr}
\crdata{ 978-1-4503-3490-7/15/11. \\
DOI: http://dx.doi.org/10.1145/2830629.2835218}

\begin{document}
\bibliographystyle{plain}
\title{Octopus: A Zero-Cost Architecture for Stream Network Monitoring}

\numberofauthors{3}
\author{
\alignauthor Andr\'es Arcia-Moret\\
\affaddr{Computer Laboratory} \\
\affaddr{University of Cambridge, UK} \\
\email{andres.arcia@cl.cam.ac.uk}
\alignauthor Jes\'us G\'omez\\
\affaddr{RESIDE} \\
\affaddr{University of Los Andes}\\
\affaddr{Venezuela}\\
\email{jagomez@ula.ve}
\alignauthor Arjuna Sathiaseelan\\
\affaddr{Computer Laboratory} \\
\affaddr{University of Cambridge, UK} \\
\email{arjuna.sathiaseelan@cl.cam.ac.uk}
}

\maketitle
%


\section{Introduction}

Considering the growing demand and popularity of Do-It-Yourself (DIY) networks,  low-cost devices managed by the people, for the people and the ease of deployment of localised/decentralised Internet services, it is mandatory for such networks to have an efficient low cost monitoring platform. Appropriate monitoring is crucial to ensure availability, responsiveness and users' Quality of Experience (QoE). This is specially relevant for the developing world where Community Networks (CN) are increasing in popularity and complexity \cite{gaia}.

CN have the capability to grow into complex arrays of interconnected nodes with different link layer technologies (wireless being more common), along with data flowing from and to not just standard Internet services but also local services: local institutional repositories, local mail servers, community clouds\cite{guifi}. Furthermore, CN users can deploy their own services catered to the local population with almost no restrictions.

Keeping track of the various network metrics in real time and to compute relevant analytics, require stream processing capabilities. Regular characteristics such as congestion, packet losses, routing inefficiencies, latency or bandwidth usage are difficult to understand and deploy from the perspective of well-known (and resource hungry) applications. It is also well-known that substantial improvements can be achieved on service performance if the monitoring information is available to the system administrator in real time. 

We propose a system that focuses on the ease of data visualization and the use of commodity hardware allowing lay people to understand the rough behaviour of their networks. Effective data visualization can also help in communicating system performance through the use of various graphical (and intuitive) elements\footnote{\scriptsize  http://www.highcharts.com}, allowing administrators to easily discover patterns and draw conclusions on system performance. 

Current emerging approaches for on-demand computation such as stream computing\footnote{\scriptsize https://en.wikipedia.org/wiki/Stream\_(computing)} are part of our motivation, as it plays an important role in small data applications.  Moreover, it addresses the problem of requesting live information that otherwise should be handled in batches. In our specific case, a stream computing approach allows the monitoring-user (with enough capabilities) to schedule the visualisation of results whenever the CPU usage (of the service holder) is low, thus assuring appropriate use of low-cost resources.


\section{Octopus}
\label{octopus-architecture}

\begin{figure}
\centering
\includegraphics[scale=0.48]{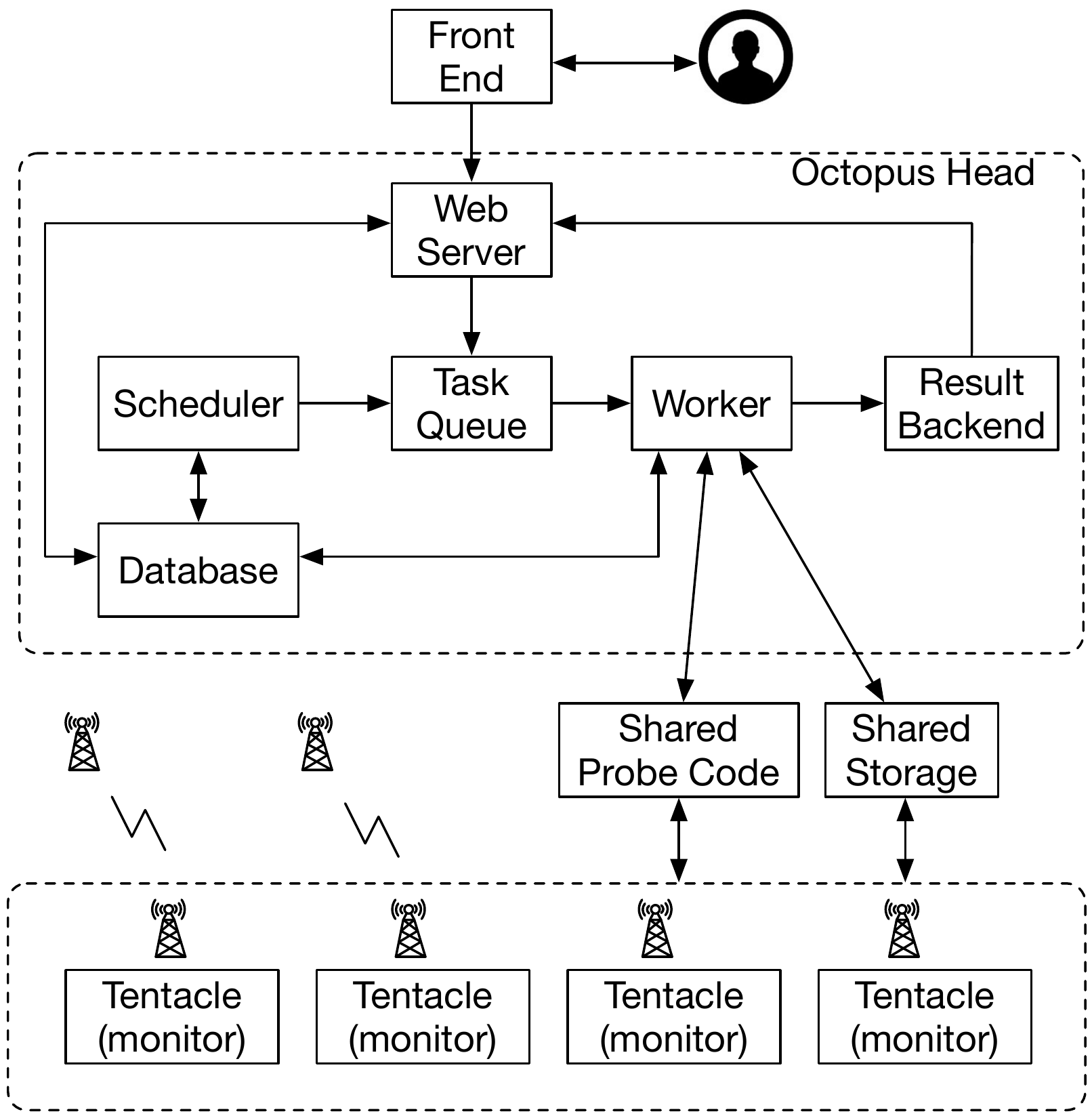}
\caption{Octopus monitor architecture}
\label{fig:octopus-architecture}
\end{figure}

Octopus Monitor (OM)\footnote{\scriptsize Available at: http://150.185.138.59/octopusmonitor} is a service to provide low-cost network monitoring solutions and data visualization to make analysis about the state of art network set-ups. Fig.~\ref{fig:octopus-architecture} summarizes the architecture of OM, a stream processing oriented monitoring system that consists of the following: (a) \textit{A central monitor entity} (the Octopus Head, OH) that controls the monitoring of different independent link and service monitors. \textbf{Octopus Head} is the entry point of the web app, it allows the interaction with users of type: \textbf{Admin} who is able to create tests and visualizations as well as manage other users, \textbf{Monitoring Users}  that own monitors and are gathering data and viewing results and, \textbf{Unauthenticated users}  that are able to view the home page, register and view the results shared by monitoring users. Heavy io/processor bound tasks are delegated to a distributed processing system. (b) \textit{Distributed monitoring elements} referred as the \textit{tentacles}. \textbf{Tentacles} are remote network monitors deployed by the monitoring users, as previously mentioned, tentacles execute periodic tests according to a schedule defined by Admin. Within tentacles, a monitoring-user specify the links that are to be monitored and the parametrized tests. A tentacle is able to load test modules and schedule them for execution on run time; tests can be executed simultaneously and exceptions are independently handled so that eventual problems go unnoticed by other concurrent tests.

In addition, the Octopus also has (c) A data repository (most likely in the cloud) that conveniently stores traces and offers the possibility of increasing the granularity of observations,  (d) A code repository to store the logic of probes, 
(e) Using the stream processing principle for retrieving pertinent information at the right time with a pertinent (appropriate) use of limited resources and (f) Use of caches to store previous expensive (heavy io/processor bound) visualizations.

\textbf{Distributed Task Processing} As the OH has to be offloaded of long running tasks, we use a distributed task processing system where one or more workers execute asynchronous tasks, and so we use task distribution system and a task queue manager (i.e., Celery/Redis). Other companion components correspond to: a \textbf{Database} for storing processed traces in compact format, user preferences and monitoring configurations. It also serves the test integration framework and the scheduler. \textbf{File Cache} to speed up previously calculated plots and avoid repetition of intensive tasks. \textbf{Static Media Server} for ready-to-consume files stored in the file cache that rarely change. Serving these files through the main web server is known to be grossly inefficient, and can overload the server unnecessarily.


\textbf{On network monitoring characteristics.} There is a wealth of innovative solutions and web applications that provide different approaches for network monitoring and benchmarking\footnote{\scriptsize http://netalyzr.icsi.berkeley.edu/, http://projectbismark.net/}. Although all of them offer roughly the same standard monitoring metrics for networks from an end-to-end perspective, none of them offer an architecture for distributed continuous monitoring within a DIY network, taking into account low-cost hardware. Metrics such as RTT, per hop RTT, packet losses, throughput, etc.  are common on active projects on network monitoring and could be independently implemented in Octopus.

\textbf{Automating network probes.} 
New probes can be part of Octopus monitoring workflow in two stages: development and deployment. At development stage, the monitoring-user defines probes, tests and visualizations; then the monitoring-user pushes the probe code into the code-repository. In order to easily deploy and integrate new probes, a simple pull from the code-repository on the OH and OT suffices.

\begin{figure}
\includegraphics[scale=0.53]{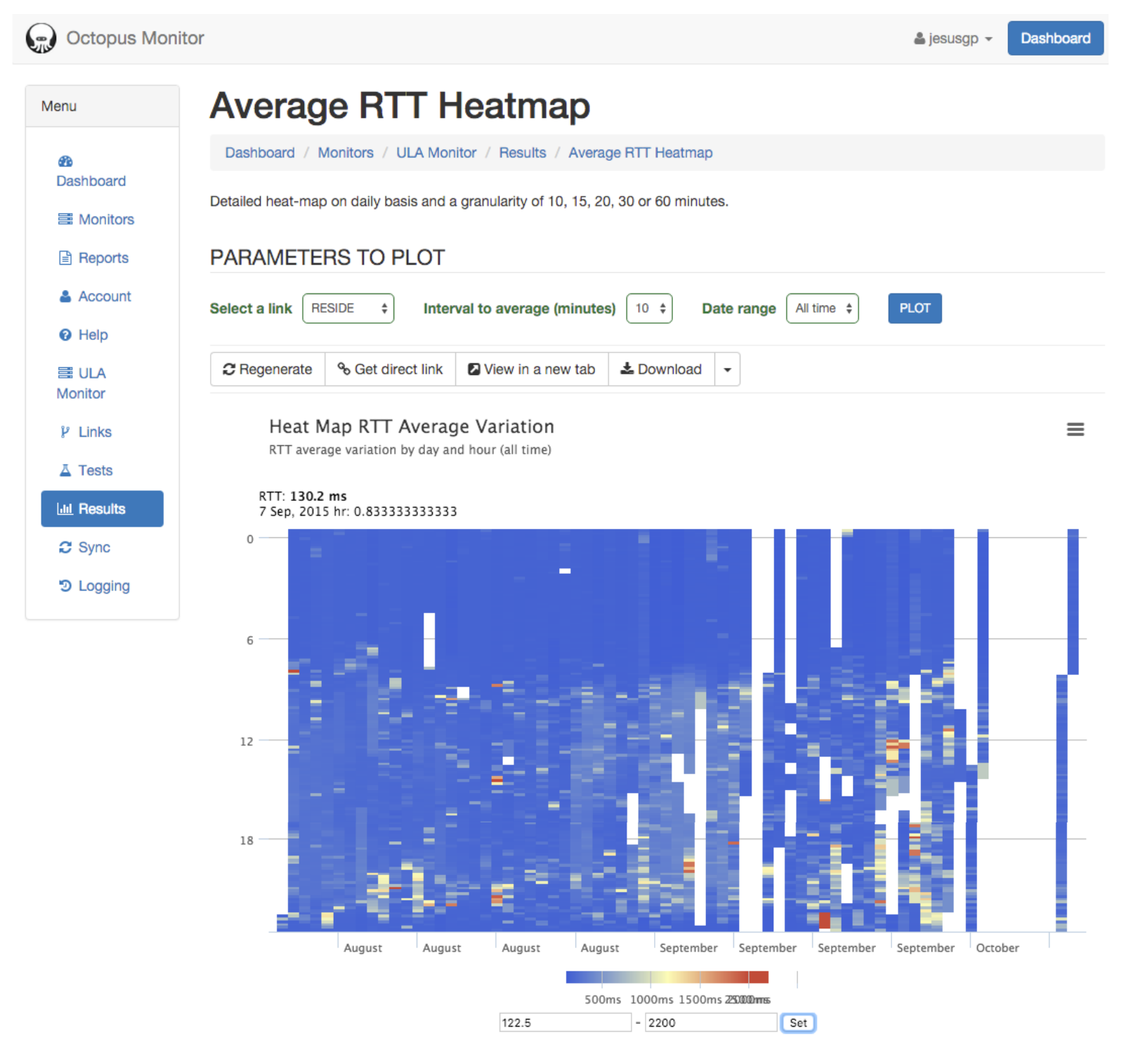}
\caption{Screen shot of Octopus, monitoring a University of Los Andes independent user service.}
\label{fig:octopus-screenshot}
\end{figure}

Finally, in Fig.~\ref{fig:octopus-screenshot} shows a screen shot of the Octopus up-and-running prototype. It shows a 2.5 month monitoring of a web-server summarizing, in a heatmap, RTTs in 10 min chunks.

\section{Conclusions}

We have developed Octopus, a prototype for low-cost network monitoring that is easily scalable and highly configurable. We can incorporate new probes and tests through a simple strategy. Octopus allows monitoring of DIY and CN for independent users. Currently we are deploying Octopus in various CNs in developing countries.

\section{Acknowledgements}
The research leading to these results has received funding from the European Union\textquotesingle s (EU) Horizon 2020 research and innovation programme under grant agreement No. 644663. Action full title: architectuRe for an Internet For Everybody, Action Acronym: RIFE. We would also like to thank Juan Tirado for comments on an earlier draft.

\end{document}